\RequirePackage{etoolbox}
\csdef{input@path}{{style/}{graphics/}}
\documentclass[MSNbibl,nameyear,seceqn,dvips]{arxstspdf}
\usepackage{graphicx}
\usepackage{url,breakurl}
\usepackage{flushend}
\usepackage{stfloats}

\volume{29}
\issue{4}
\pubyear{2014}
\firstpage{579}
\lastpage{595}
\doi{10.1214/14-STS486} % kopijuoti is 'New paper accepted'
\docsubty{FLA}

\makeatletter
 \fnbelowfloat
\newproclaim{example}{Example}
\newproclaim{definition}{Definition}
\newtheorem{theorem}{Theorem}
\newtheorem{corollary}{Corollary}
\newproclaim{Rule}{Rule}
\newproclaim{Remark}{Remark}
\newcommand{\compos}{\mbox{$\bot\!\!\!\!\bot$}}

\def\ci{\mbox{$\perp\!\!\!\!\perp$}}

\renewcommand{\emptyset}{\varnothing}
\newcommand{\po}{\operatorname{do}}
\newcommand{\pa}{\operatorname{pa}}
\makeatother

\begin{document}
\begin{frontmatter}

\title{External Validity: From Do-Calculus to Transportability Across Populations}
% kai straipsnis turi susijusiu diskusiju ir rejoinder'iu
%rejoinder at \relateddoi{r}{10.1214/00-STSXXXX}.}
\runtitle{External Validity: From do-calculus to Transportability}

\begin{aug}
\author[A]{\fnms{Judea} \snm{Pearl}\corref{}\ead[label=e1]{judea@cs.ucla.edu}}
\and
\author[A]{\fnms{Elias} \snm{Bareinboim}\ead[label=e2]{eb@cs.ucla.edu}}
\runauthor{J. Pearl and E. Bareinboim}

\affiliation{University of California, Los Angeles}

\address[A]{Judea Pearl is Professor and Elias Bareinboim is Ph.D. Candidate,
Computer Science Department,
University of California,
Los Angeles, California 90095-1596, USA \printead{e1,e2}.}
\end{aug}

% ABSTRACT
%
\begin{abstract}
The generalizability of empirical findings to new
environments, settings or populations, often called
``external validity,'' is essential in most scientific explorations.
This paper treats a particular problem of generalizability,
called ``transportability,'' defined
as a license to transfer causal
effects learned in experimental studies to
a new population, in which only observational studies
can be conducted. We introduce a formal representation
called ``selection diagrams'' for expressing
knowledge about differences and commonalities between
populations of interest and, using this representation, we
reduce questions of transportability to
symbolic derivations in the do-calculus. This reduction
yields graph-based procedures for deciding, prior to observing any
data, whether causal effects in
the target population can be inferred
from experimental findings in the study population. When
the answer is affirmative, the procedures identify what
experimental and observational findings need be obtained
from the two populations, and how they can be combined
to ensure bias-free transport.
\end{abstract}

% KEYWORDS
\begin{keyword}
\kwd{Experimental design}
\kwd{generalizability}
\kwd{causal effects}
\kwd{external validity}
\end{keyword}
\end{frontmatter}

%s1 #&#
\section{Introduction: Threats vs. Assumptions} \label{sec1}

Science is about generalization, and generalization
requires that conclusions obtained in the laboratory
be transported and applied elsewhere, in an environment
that differs in many aspects from that of the laboratory.

Clearly, if the target environment is
arbitrary, or drastically different
from the study environment nothing can be transferred and
scientific progress
will come to a standstill. However, the fact that most
studies are conducted with
the intention of applying the results elsewhere means that
we usually deem the
target environment sufficiently similar to the study
environment to justify the
transport of experimental results or their ramifications.
% use later, in section 3
%studies requires
%some understanding of
%the reasons for the differences{\chr"E2}{\chr"80}{\chr"9D} (page 11).

Remarkably, the conditions that permit such transport have not received
systematic formal treatment.
In statistical practice, problems related to
combining and generalizing from diverse studies
are handled by methods of meta analysis
(\cite{glass:1976}; \cite{hedges:etal85}; \cite{owen:2009}),
or hierarchical models (\cite{gelmanhill07}),
%in which results of diverse studies are pooled together by weighted
%averages, and performance is evaluated primarily by simulation.
in which results of diverse studies are pooled together by standard
statistical procedures (e.g., inverse-variance reweighting in
meta-analysis, partial pooling in hierarchical modeling) and rarely
make explicit distinction between experimental and observational
regimes; performance is evaluated primarily by simulation.

To supplement these methodologies, our
paper provides theoretical guidance in the
form of limits on what can be achieved
in practice, what problems are likely to be encountered when
populations differ significantly from each other,
what population differences can be circumvented by
clever design and what differences constitute theoretical
impediments, prohibiting generalization by any means whatsoever.

On the theoretical front, the standard literature on this topic,
falling under rubrics such as ``external validity'' (\cite{campbell:sta63,manski:07}), ``heterogeneity'' (\cite{hofler:etal10}),
``quasi-experiments'' (\cite{shadish:etal01}, Chapter~3; \cite{adelman:1991}),\footnote{\citet{manski:07} defines ``external
validity'' as follows: ``An experiment is said to have ``external
validity'' if the distribution of outcomes realized by a treatment
group is the same as the distribution of outcome that would be realized
in an actual program.''
\citet{campbell:sta63}, page 5, take a slightly broader view:
````External validity'' asks the question of generalizability: to what
populations, settings, treatment variables, and measurement variables
can this effect be generalized?''\label{foot1}}
consists primarily of ``threats,'' namely, explanations of what may go
wrong when we
try to transport results from one study to
another while ignoring their differences.
Rarely do we find an
analysis of ``licensing assumptions,'' namely, formal
conditions under which the transport of results across differing
environments or populations
is licensed from first principles.\footnote{\citet{hernan:van11}
%%%Hernan and VanderWeele (2011)
studied such conditions in the context of compound treatments, where
we seek to predict the effect of one version of a treatment
from experiments with a different version. Their
analysis is a special case of the theory developed
in this paper (\cite{petersen:11}). A related application is reported
in \citet{robinsetal:08} where a treatment strategy is extrapolated
between two biological similar populations under different
observational regimes. }

The reasons for this asymmetry are several. First, threats
are safer to cite than
assumptions. He who cites ``threats'' appears prudent,
cautious and thoughtful,
whereas he who seeks licensing assumptions risks suspicions of
attempting to endorse those assumptions.

Second, assumptions are self-destructive in their honesty.
The more explicit
the assumption, the more criticism it invites, for it tends
to trigger a richer space
of alternative scenarios in which the assumption may fail.
Researchers prefer
therefore to declare threats in public and make assumptions
in private.

Third, whereas threats can be communicated in plain English,
supported by
anecdotal pointers to familiar experiences, assumptions
require a formal language
within which the notion ``environment'' (or
``population'') is given precise characterization,
and differences among environments can be encoded and
analyzed.

%The advent of causal diagrams (wright, heise, davis, spirtes93,
%pearl95) \citet{pearl 95,greenland:etal99a,spirtes:etal00,pearl:09}
%provides such a language and renders the formalization of
%transportability possible.

The advent of causal diagrams (\cite{wright:21}; \cite{heise:75}; \cite{davis84};
\cite{verma:pea88}; \cite{spirtes:etal93}; \cite{pearl:95}) together with models of
interventions (\cite{haavelmo:43}; \cite{strotz:wol60}) and counterfactuals
(\cite{neyman:23a}; \cite{rubin:74}; \cite{robins:86}; \cite{balke:pea95}) provides such a
language and renders the formalization of transportability possible.

Armed with this language, this
paper departs from the
tradition of communicating ``threats'' and embarks
instead on the
task of formulating ``licenses to transport,'' namely,
assumptions that, if they
held true, would permit us to transport results across
studies.

In addition, the paper uses the inferential
machinery of the do-calculus (\cite
{pearl:95}; \cite{Koller+Friedman:09}; \cite{huang:val06}; \cite{shpitser:pea06-r329}) to derive
algorithms
for deciding whether transportability is feasible
and how experimental and observational
findings can be combined to yield unbiased estimates
of causal effects in the target population.

The paper is organized as follows. In Section~\ref{sec2a}, we review
the foundations of structural
equations modeling (SEM), the question of identifiability
and the do-calculus that emerges from these foundations.
(This section
can be skipped by readers familiar with
these concepts and tools.)
In Section~\ref{sec2}, we motivate the question of transportability
through simple examples, and illustrate how the
solution depends on the causal story behind the problem.
In Section~\ref{sec3}, we formally define the notion of transportability
and reduce it to a problem of symbolic transformations
in do-calculus. In Section~\ref{sec4}, we provide a graphical
criterion for deciding transportability and estimating
transported causal effects.
We conclude in Section~\ref{sec8} with brief discussions of related
problems of external validity, these include
statistical transportability, and meta-analysis.

%s2 #&#
\section{Preliminaries: The Logical Foundations of Causal Inference}
\label{sec2a}

The tools presented in this paper were developed in the context of
nonparametric Structural Equations Models (SEM), which is one among
several approaches to causal inference, and goes back to (\cite
{haavelmo:43}; \cite{strotz:wol60}). Other approaches include, for example,
potential-outcomes (\cite{rubin:74}), Structured Tree Graphs (\cite
{robins:86}), decision analytic (\cite{dawid:02}), Causal Bayesian
Networks (\cite{spirtes:etal00}; \cite{pearl:2k}, Chapter~1; \cite{bcp:12}), and Settable Systems (\cite{white:09}). We will first
describe the generic features common to all such approaches, and then
summarize how these features are represented in SEM.\footnote{ We use
the acronym SEM for both parametric and nonparametric representations
though, historically, SEM practitioners preferred the former (\cite
{bollen-pearl:12}). \citet{pearl:11} has used the term Structural Causal
Models (SCM) to eliminate this confusion. While comparisons of the
various approaches lie beyond the scope of this paper, we nevertheless
propose that their merits be judged by the extent to which each
facilitates the functions described below.}

%s2.1 #&#
\subsection{Causal Models as Inference Engines} \label{sec:21}

From a logical viewpoint, causal analysis relies on
causal assumptions that cannot be deduced from
(nonexperimental) data. Thus, every approach
to causal inference must provide a
systematic way of encoding, testing and combining
these assumptions with data. Accordingly, we view causal modeling
as an inference engine that takes three inputs and produces
three outputs.
The inputs are:
\begin{enumerate}[I-3.]
\item[I-1.] A set $A$ of qualitative causal \emph{assumptions}
which the investigator is prepared to defend on
scientific grounds, and a model $M_A$ that encodes these assumptions
mathematically. (In SEM, $M_A$ takes the form
of a diagram or a set of unspecified functions. A typical
assumption is that no direct effect exists between a pair of variables
(known as exclusion restriction),
or that an omitted factor, represented by
an error term, is independent of other such factors observed or
unobserved, known as well as unknown.
\item[I-2.] A set $Q$ of \emph{queries} concerning causal or
counterfactual relationships among variables of interest.
%%Insert-9 (after "variables of interest.")
In linear SEM, $Q$ concerned the magnitudes of structural coefficients
but, in general, $Q$ may address causal relations directly, for example:
\begin{enumerate}[$Q_2$:]
\item[$Q_1$:] What is the effect of treatment $X$ on outcome $Y$?

\item[$Q_2$:] Is this employer practicing gender discrimination?
\end{enumerate}
In principle, each query $Q_i \in Q$ should be ``well defined,'' that
is, computable from any fully specified model $M$ compatible with $A$.
(See Definition~\ref{def11} for formal characterization of a model,
and also Section~\ref{sec:id} for the problem of identification in
partially specified models.)
\item[I-3.] A set $D$ of experimental or non-experimental \emph
{data}, governed by a joint probability distribution presumably
consistent with A.
\end{enumerate}
%
%f1 #&#
\begin{figure*}

\includegraphics{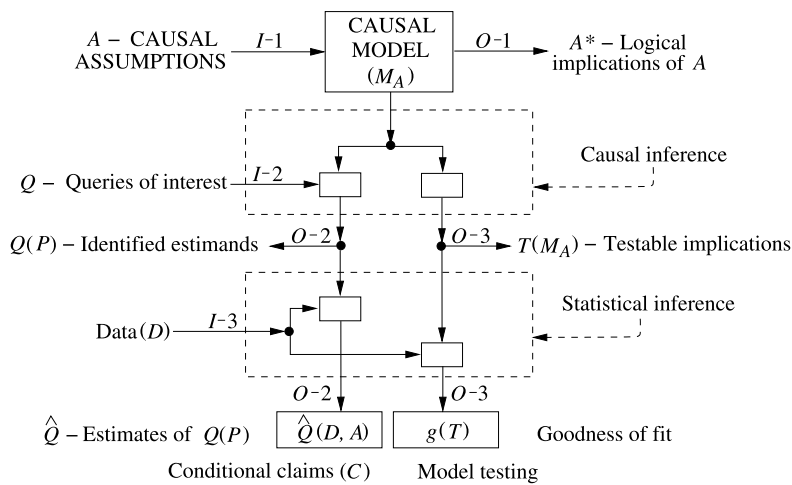}

\caption{Causal analysis depicted as an inference engine
converting assumptions $(A)$, queries $(Q)$, and data $(D)$
into logical implications $(A^*)$, conditional claims $(C)$,
and data-fitness indices $(g(T))$.}\label{atlanta-slide1}
\end{figure*}

The outputs are:
\begin{enumerate}[O-3.]
\item[O-1.] A set $A^*$ of statements which are the logical
implications of $A$, %Insert-9.1 (after "inplications of A")
separate from the data at hand. For example, that $X$ has
no effect on $Y$ if we hold $Z$ constant, or that
$Z$ is an instrument relative to \{$X$, $Y$\}.

\item[O-2.] %%insert-10 Replace item O-2 with:
A set $C$ of data-dependent \emph{claims} concerning
the magnitudes or likelihoods of the target queries
in $Q$, each contingent on $A$.
$C$ may contain, for example, the estimated mean and variance
of a given structural parameter, or the expected effect of
a given intervention. Auxiliary to $C$, a causal model should also yield
an estimand $Q_i(P)$ for each query in $Q$, or a determination
that $Q_i$ is not identifiable from $P$ (Definition~\ref
{def-identifiability}). %%1.)
\item[O-3.] A list $T$ of testable statistical implications of $A$
(which may or may not be part of O-2),
and the degree $g(T_i), T_i \in T$, to which the data agrees with each
of those
implications. A typical
implication would be a conditional independence assertion,
or an equality constraint between two probabilistic
expressions.
Testable constraints should be read from the model $M_A$ (see
Definition~\ref{sec3-def}),
and used to confirm or disconfirm the model against the data.
\end{enumerate}
The structure of this inferential exercise is shown schematically in
Figure~\ref{atlanta-slide1}. For a comprehensive review on
methodological issues, see \citeauthor{pearl:09-r350} (\citeyear{pearl:09-r350}, \citeyear{pearl:12-r370}).

%f1 ###
% \psfig{figure=figure1-r355.eps,width=5.25in}
%associated diagrams, showing (a) independent unobserved exogenous
%variables
%(connected by dashed arrows), (b) dependent exogenous variables, and
%(c) an equivalent, more traditional notation, in which latent
%variables are enclosed in ovals.\label{r355-fig1}}

%% section 3.2 from r355
%s2.2 #&#
\subsection{Assumptions in Nonparametric Models}
\label{ss3.2}

%of identification was undertaken through the structural formalism as
%the underlying data-generating model, all the results regarding
%interventions hold in a similar fashion over any formalism equivalent
%to Semi-Markovian models, which is precisely the second layer of the
%causal hierarchy discussed in \cite{pearl:2k}. In other words, the
%structural equation assumptions just gives us a more cognitive
%meaningful way for judging assumptions (and it entails the second
%layer), but remarkably, it does not preclude any result when one
%decided to undertake the analysis only through a purely
%experimentalist perspective.

%
A structural equation model (SEM) $M$ is defined as follows.

%
%de1 #&#
\begin{definition}[(Structural equation model
(\cite{pearl:2k}, page 203))]\label{def11}
\begin{enumerate}[4.]
\item[1.] A set $U$ of background or exogenous variables,
representing factors outside the model, which
nevertheless affect relationships within the model.
\item[2.] A set $V = \{V_1,\ldots,V_n\}$ of endogenous
variables, assumed to be observable. Each of these
variables is functionally dependent on some subset
$PA_i$ of $U \cup V$.

%f2 #&#
\begin{figure}[b]

\includegraphics{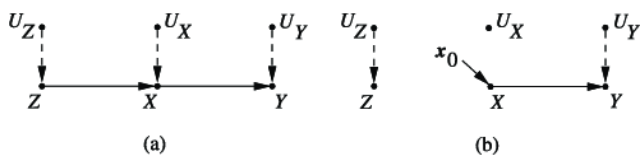}

\caption{The diagrams associated with \textup{(a)} the
structural model of equation (\protect\ref{eq5a}) and
\textup{(b)} the modified model
of equation (\protect\ref{eqx}), representing the intervention $\po(X=x_0)$.}\label{figure2-r355}
\end{figure}

\item[3.] A set $F$ of functions $\{f_1, \ldots,f_n\}$ such that each
$f_i$ determines the value of $V_i \in V$,
$v_i = f_i(\pa_i, u)$.
\item[4.] A joint probability distribution $P(u)$ over $U$.
\end{enumerate}
\end{definition}

A simple SEM model is depicted in Figure~\ref{figure2-r355}(a), which
represents
the following three functions:
%
%e2.1 #&#
\begin{eqnarray}
\label{eq5a} z& =& f_{Z}(u_Z),
\nonumber
\\
x& =& f_{X}(z,u_X),
\\
y& =& f_{Y}(x,u_Y),
\nonumber
\end{eqnarray}
where in this particular example, $U_Z$, $U_X$ and $U_Y$ are assumed
to be jointly independent but otherwise arbitrarily distributed.
Whenever dependence exists between any two exogenous variables, a
bidirected arrow will be added to the diagram to represent this
dependence (e.g., Figure~\ref{fig2}).\footnote{More precisely, the absence of bidirected arrows implies marginal
independences relative of the respective exogenous variables. In other
words, the set of all bidirected edges constitute an i-map of $P(U)$
(\cite{richardson:03}).
} %\foSotnote{Conversely, the lack of bidirected arrow indicates both
%marginal independence and conditional independence given any subset of
%the other exogenous. }
Each of these functions represents a causal process (or mechanism) that
determines the value of the left variable (output) from the values on
the right variables (inputs), and is assumed to be invariant unless
explicitly intervened on. The absence of a variable
from the right-hand side of an equation encodes the assumption that
nature ignores that variable in the process of determining the value of
the output variable.
For example, the absence of variable $Z$ from the arguments of $f_Y$
conveys the empirical claim that variations in $Z$ will leave Y
unchanged, as long as variables $U_Y$ and $X$ remain constant.

It is important to distinguish between a \textit{fully specified
model} in which $P(U)$ and the collection of functions $F$ are
specified and a \textit{partially specified model}, usually in the
form of a diagram. The former entails one and only one observational
distribution $P(V)$; the latter entails a set of observational
distributions $P(V)$ that are compatible with the graph (those that can
be generated by specifying $\langle F, P(u)\rangle$). % as well
%constraints over the set of interventional and counterfactual
%distributions.

%The former entails one and only one observational distribution $P(V)$
%together with a unique distribution for every intervention and every
%counterfactual. The latter defines a set of observational
%distributions $P(V)$ that are compatible with the graph as well
%constraints over the set of interventional and counterfactual
%distributions.

%s3.2.1 ###
%s2.3 #&#
\subsection{Representing Interventions, Counterfactuals and Causal Effects}

This feature of invariance permits us to derive powerful claims about
causal effects and
counterfactuals, even in nonparametric models, where all functions and
distributions remain unknown. This is done through a mathematical
operator called $\po(x)$, which simulates physical interventions by
deleting certain functions from the model, replacing them with a
constant $X = x$, while keeping the rest of the model unchanged (\cite
{haavelmo:43}; \cite{strotz:wol60}; \cite{pearl12-r391}).
For example, to emulate an intervention $\po(x_0)$ that sets $X$ to a
constant $x_0$ in model $M$ of Figure~\ref{figure2-r355}(a), the
equation for $x$ in equation (\ref{eq5a}) is replaced by $x = x_0$,
and we obtain a new model, $M_{x_0}$,
%e6 ###
%e2.2 #&#
\begin{eqnarray}\label{eqx}
z& =& f_{Z}(u_Z),
\nonumber
\\
x& =& x_0,
\\
y& =& f_{Y}(x,u_Y),
\nonumber
\end{eqnarray}
the graphical description of which is shown in Figure~\ref{figure2-r355}(b).

The joint distribution associated with this modified model,
denoted $P(z,y|\po(x_0))$ describes the post-intervention distribution
of variables $Y$ and $Z$
(also called ``controlled'' or ``experimental'' distribution), to be
distinguished from the preintervention distribution,
$P(x,y,z)$, associated with the original model of
equation (\ref{eq5a}).
For example, if $X$ represents a treatment variable, $Y$ a response
variable, and $Z$ some covariate that affects the amount
of treatment received, then the distribution $P(z,y|\po(x_0))$
gives the proportion of individuals that would attain
response level $Y = y$ and covariate level $Z=z$
under the hypothetical situation in which treatment $X = x_0$ is
administered uniformly to the population.\footnote{Equivalently,
$P(z,y|\po(x_0))$ can be interpreted
as the joint probability of $(Z=z, Y=y)$ under a randomized experiment
among units receiving treatment level $X=x_0$. Readers versed
in potential-outcome notations may interpret $P(y|\po(x),z)$
as the probability $P(Y_x =y | Z_x = z)$, where
$Y_x$ is the potential outcome under treatment $X=x$.}

In general, we can formally define the postintervention distribution by
the equation
%e7 ###
%e2.3 #&#
\begin{equation}\label{eq111-7}
P_M\bigl(y|\po(x)\bigr) = P_{M_x}(y). %%eq. (111)
\end{equation}
In words, in the framework of model $M$,
the postintervention distribution of outcome $Y$
is defined as the probability that model $M_x$ assigns
to each outcome level $Y=y$.
From this distribution, which is readily computed from any fully
specified model $M$, we are able to
assess treatment efficacy by comparing
aspects of this distribution at different levels of~$x_0$.\footnote
{Counterfactuals are defined similarly through the equation $ Y_x(u) =
Y_{M_x}(u)$
(see \cite{pearl:09}, Chapter~7), but will not be needed for
the discussions in this paper.}

%. For example, when only the graph is available whether we can
%identify the interventional distribution $P(y | do(x))$ which is
%defined as simply the arguments of the functions $F$ and the
%relationships between the $U$'s: given assumptions set $A$ (as
%embodied in the coarse description of the model as depicted in Fig.
%distribution, $P(y | do(x), z)$, be estimated from data collected
%under a preinterventional regime $P(z,x,y)$?

%s2.4 #&#
\subsection{Identification, d-Separation and Causal Calculus} \label{sec:id}
A central question in causal analysis is the question of \emph
{identification} of causal queries (e.g., the effect of intervention
$\po(X=x_0)$) from a combination of data and a partially specified
model, for example, when only the graph is given and neither the
functions $F$ nor the distribution of $U$. In linear parametric
settings, the question of identification reduces to asking whether some
model parameter,
$\beta$, has a unique solution in terms of the parameters
of $P$ (say the population covariance matrix).
In the nonparametric formulation, the notion of ``has a unique
solution'' does not directly
apply since quantities such as $Q(M) = P(y|\po(x))$ have
no parametric signature and are defined
procedurally by simulating an intervention in a causal
model $M$, as in equation (\ref{eqx}). The following definition
captures the requirement that $Q$ be estimable from the data:
%

%de2 #&#
\begin{definition}[(Identifiability)]\label{def-identifiability}
%definition which deals with the identifiability of the parameter set $
%In our case, one should think about the query $Q=P(y|do(x))$ as a
%function $Q = g(\theta)$ where $\theta$ is the pair $F \cup P(u)$ that
%characterizes a fully specified model $M$. }
%definition which deals with the identifiability of the parameter set $
%In our case, one should think about the query $Q=P(y|do(x))$ as a
%function $Q = g(\theta)$ where $\theta$ is the pair $F \cup P(u)$ that
%characterizes a fully specified model $M$. }
A causal query $Q(M)$ is
identifiable, given a set of assumptions $A$, if for any
two (fully specified) models, $M_1$ and $M_2$, that satisfy $A$, we
have \footnote{An implication similar to (\ref{eq00}) is used in the
standard statistical definition of parameter identification, where it
conveys the uniqueness of a parameter set $\theta$ given a
distribution $P_{\theta}$ (\cite{lehmann-casella:98}). To see the
connection, one should think about the query $Q=P(y|\po(x))$ as a
function $Q = g(\theta)$ where $\theta$ is the pair $F \cup P(u)$
that characterizes a fully specified model $M$.}
%
%e10 ###
%e2.4 #&#
\begin{equation}\label{eq00}
P(M_1) = P(M_2) \Rightarrow Q(M_1) =
Q(M_2). %%eq. (00)
\end{equation}
\end{definition}

In words, the functional details of $M_1$ and $M_2$ do not matter;
what matters is that the assumptions in $A$ (e.g., those encoded in the
diagram) would
constrain the variability of those details in such
a way that equality of $P$'s would entail equality of $Q$'s.
When this happens, $Q$ depends on $P$ only, and should therefore
be expressible in terms of the parameters of $P$.

When a query $Q$ is given in the form of a do-expression,
for example, $Q=P(y|\po(x), z)$, its identifiability can
be decided systematically using an algebraic procedure
known as the do-calculus (\cite{pearl:95}).
It consists of three inference rules that permit us to map
interventional and observational distributions whenever
certain conditions hold in the causal diagram~$G$.

The conditions that permit the application these inference rules
can be read off the diagrams using a graphical criterion
known as d-separation (\cite{pearl:88a}).

%de3 #&#
\begin{definition}[(d-separation)] \label{sec3-def}
A set $S$ of nodes is said to block a path $p$ if either
\begin{enumerate}[2.]
\item$p$ contains at least one arrow-emitting node that is in $S$, or
\item $p$ contains at least one collision node that is outside $S$ and
has no descendant in $S$.
\end{enumerate}
If $S$ blocks \emph{all} paths from set $X$ to set $Y$, it is said to
``d-separate $X$ and $Y$,''
and then, it can be shown that variables $X$ and $Y$ are independent given
$S$, written $X \compos Y|S$.\footnote{See \citet{hayduk:etal03},
\citet{glymour:gre08} and \citet{pearl:09}, page 335, for a gentle
introduction to d-separation.}
\end{definition}

D-separation reflects conditional independencies that hold in any
distribution $P(v)$ that is compatible
with the causal assumptions $A$ embedded in the diagram. To illustrate,
the path $U_Z \rightarrow Z \rightarrow X \rightarrow Y$ in Figure~\ref{figure2-r355}(a) is blocked by $S=\{Z\}$ and by $S=\{X\}$, since each
emits an arrow
along that path. Consequently, we can infer that the conditional
independencies $U_Z \compos Y|Z$ and $U_Z \compos Y|X$ will be
satisfied in any
probability function that this model can generate,
regardless of how we parameterize the arrows.
Likewise, the path $U_Z \rightarrow Z \rightarrow X \leftarrow U_X$ is blocked
by the null set $\{\emptyset\}$, but it is not blocked by $S = \{Y\}$
since $Y$ is a descendant of the collision node $X$. Consequently,
the marginal independence $U_Z \compos U_X$ will hold in
the distribution, but $U_Z \compos U_X|Y$ may or may not hold.\footnote
{This special handling of collision nodes (or \emph{colliders},
e.g., $Z \rightarrow X\leftarrow U_X$)
reflects a general phenomenon known as
\emph{Berkson's paradox} (\cite{berkson:46}),
whereby observations on
a common consequence of two independent causes
render those causes dependent. For example, the outcomes of
two independent coins are rendered dependent by the testimony
that at least one of them is a tail.}

%s2.5 #&#
\subsection{The Rules of do-Calculus}
Let $X$, $Y$, $Z$ and $W$ be arbitrary disjoint sets of nodes in a
causal DAG $G$. We denote by $G_{\overline{X}}$ the graph obtained
by deleting from $G$ all arrows pointing to nodes in $X$.
Likewise, we denote by $G_{\underline{X}}$ the graph obtained by
deleting from $G$ all arrows emerging from nodes in $X$. To represent
the deletion of both incoming and outgoing arrows,
we use the notation $G_{\overline{X}\underline{Z}}$.\vspace*{2pt}
%
%f3 #&#
\begin{figure*}

\includegraphics{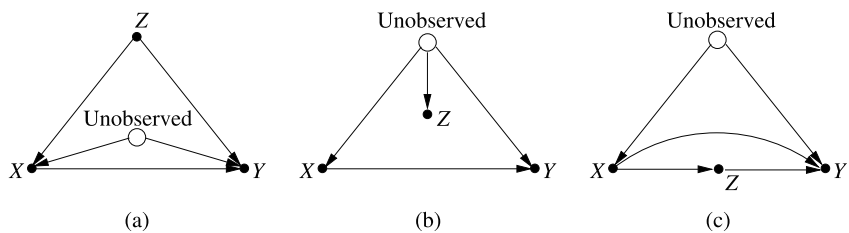}

\caption{Causal diagrams depicting Examples \protect\ref{ex1}--\protect\ref{ex3}.
In \textup{(a)} $Z$ represents ``age.'' In \textup{(b)}, $Z$ represents
``linguistic skills'' while age (in hollow circle) is
unmeasured. In \textup{(c)}, $Z$ represents a biological marker
situated between the treatment $(X)$ and a disease~$(Y)$.}
\label{fig1}
\end{figure*}

The following three rules are valid for every interventional
distribution compatible with $G$:

\begin{Rule}[(Insertion/deletion of observations)]\label{r1}
%
%e2.5 #&#
\begin{eqnarray}
&&P\bigl(y | \po(x), z, w\bigr)\nonumber \\[-8pt]\\[-8pt]
&&\quad = P\bigl(y | \po(x), w\bigr) \quad \mbox{if } (Y \ci Z
| X, W)_{G_{\overline{X}}}.\nonumber
\end{eqnarray}
\end{Rule}

\begin{Rule}[(Action/observation exchange)]\label{r2}
%
%e2.6 #&#
\begin{eqnarray}
&&\hspace*{10pt}P\bigl(y | \po(x), \po(z), w\bigr)\nonumber\\[-8pt]\\[-8pt]
&&\hspace*{10pt}\quad  = P\bigl(y | \po(x), z, w\bigr) \quad \mbox{if }
(Y \ci Z | X, W)_{G_{\overline{X} \underline{Z} }}.\nonumber
\end{eqnarray}
\end{Rule}

\begin{Rule}[(Insertion/deletion of actions)]\label{r3}
%
%e2.7 #&#
\begin{eqnarray}
&&\hspace*{10pt}P\bigl(y | \po(x), \po(z), w\bigr) \nonumber\\[-8pt]\\[-8pt]
&&\hspace*{10pt}\quad = P\bigl(y | \po(x), w\bigr) \quad \mbox{if } (Y
\ci Z | X, W)_{G_{\overline{X} \overline{Z(W)} }},\nonumber
\end{eqnarray}
where $Z(W)$ is the set of $Z$-nodes that are not ancestors of any
$W$-node in $G_{\overline{X}}$.
%%\end{theorem}
\end{Rule}

To establish identifiability of a query $Q$, one needs to repeatedly
apply the rules of do-calculus to $Q$,
until the final expression no longer contains a
do-operator;\footnote{Such derivations are illustrated in graphical
details in \citet{pearl:09}, page 87.} this renders it estimable from
nonexperimental data. The do-calculus was proven to be complete for
the identifiability of causal effects in the form $Q = P(y | \po(x), z)$
(\cite{shpitser:pea06-r329}; \cite{huang:val06}),
which means that if $Q$ cannot be expressed in terms of the probability
of observables $P$ by repeated application of these three rules, such
an expression does not exist. In other words, the query is not
estimable from observational studies without making further
assumptions, for example, linearity, monotonicity, additivity, absence
of interactions, etc.

% which means that if an equality cannot be established
%by repeated application of these three rules, this equality cannot be
%obtained by any other method.

We shall see that, to establish transportability, the
goal will be different; instead of eliminating do-operators from the
query expression,
we will need to separate them from a set of variables S
that represent disparities between populations.

%Remarkably, the problem of identification of causal effects is part of
%the \textit{internal validity} of the experimental research in
%question. Other issues that may arise when evaluating internal
%validity is the problem of \textit{selection bias}, which was given
%formal treatment for general scenarios under mild set of assumptions
%in \citet{bp:12a}. In regard to the problem of transportability
%(external validity), we similarly build on the language of structural
%equations and the $do$-calculus, but in a fundamentally different way;
%we no longer assume that there is only one population of interest in
%our model of universe, but we explicit acknowledge in the model that
%different populations might exist, and further, that they also can be
%related. This simple acknowledgement implies that the relationship
%between these populations have to be used to determine the external
%validity of a certain causal claim. In the philosophical perspective,
%this complements the semantics of causal diagrams, and also requires
%new formal treatment, which is the subject of the remaining of this
%paper.

%s3 #&#
\section{Inference Across Populations: Motivating Examples} \label{sec2}

To motivate the treatment of Section~\ref{sec3}, we first demonstrate
some of the subtle questions that transportability entails through
three simple examples, informally depicted in Figure~\ref{fig1}.

%
%ex1 #&#
\begin{example}\label{ex1} %%Example-1
Consider the graph in Figure~\ref{fig1}(a) that represents
cause-effect relationships in the pretreatment population in Los
Angeles. We conduct a randomized trial in Los Angeles and estimate the
causal effect of exposure $X$ on
outcome $Y$ for\vspace{2pt} every age group $Z=z$.\footnote{Throughout the paper,
each graph represents the causal structure of the population prior to
the treatment, hence $X$ stands for the level of treatment taken by an
individual out of free choice.}\tsup{,}\footnote{The arrow from $Z$ to $X$
represents the tendency of older people to seek treatment more often
than younger people, and the arrow from $Z$ to $Y$ represents the
effect of age on the outcome.} We now wish to
generalize the results to the population of New York City (NYC),
but data alert us to the fact that the study
distribution $P(x,y,z)$ in LA is significantly different
from the one in NYC (call the latter $P^*(x,y,z)$).
In particular, we notice that the average
age in NYC is significantly higher than that in LA.
How are we to estimate the causal effect
of $X$ on $Y$ in NYC, denoted $P^*(y|\po(x))$?
\end{example}

Our natural inclination would be to assume that age-specific
effects are invariant across cities and so, if the LA study
provides us with (estimates of) age-specific causal effects
$P(y|\po(x), Z=z)$, the overall causal effect in NYC should be
%
%e3.1 #&#
\begin{equation}\label{eq1}
P^*\bigl(y|\po(x)\bigr) = \sum_z P\bigl(y|
\po(x), z\bigr) P^*(z). %% Eq. (1)
\end{equation}

This {\emph{transport formula}} combines experimental results
obtained in LA, $P(y|\po(x), z)$, with observational
aspects of NYC population, $P^*(z)$, to obtain an
experimental claim $P^*(y|\po(x))$ about NYC.\footnote{At first glance,
equation (\ref{eq1}) may be regarded as
a routine application of ``standardization'' or ``recalibration''---a statistical extrapolation method that can be traced back to
a century-old tradition in demography and political
arithmetic (\cite{westergaard:1916}; \cite{yule:1934}; \cite{lane:nel82}).
%%%(Westergaard, 1916; Yule, 1937)
On a second thought, it raises the deeper question of
why we consider age-specific effects to be invariant
across populations. See discussion following Example~\ref{ex2}.}

Our first task in this paper will be to explicate the assumptions
that renders this extrapolation valid. We ask,
for example, what must we assume about other confounding variables
beside age, both latent and observed, for equation (\ref{eq1})
to be valid, or, would the same transport formula
hold if $Z$ was not age, but some proxy for age, say,
language proficiency. More intricate yet, what
if $Z$ stood for an exposure-dependent variable,
say hyper-tension level, that stands between $X$
and $Y$?

Let us examine the proxy issue first.

%
%ex2 #&#
\begin{example}\label{ex2}
 %%Example~2
Let the variable $Z$ in Example~\ref{ex1} stand for
subjects language proficiency, and let us assume that
$Z$ does not affect exposure $(X)$ or outcome $(Y)$, yet it correlates
with both,
being a proxy for age which is not measured
in either study [see Figure~\ref{fig1}(b)]. Given the observed
disparity $P(z)\neq P^*(z)$,
how are we to estimate the causal effect
$P^*(y|\po(x))$ for the target population of NYC from the
$z$-specific causal effect $P(y|\po(x),z)$ estimated
at the study population of LA?
\end{example}

The inequality $P(z) \neq P^*(z)$ in this example may reflect either
age difference
or differences in the
way that $Z$ correlates with age.
If the two cities enjoy identical age
distributions and NYC residents acquire
linguistic skills at a younger age,
then since $Z$ has no effect whatsoever on $X$ and $Y$,
the inequality $P(z) \neq P^*(z)$ can be ignored and, intuitively,
the proper transport formula would be
%
%e3.2 #&#
\begin{equation}\label{eq2}
P^*\bigl(y|\po(x)\bigr) = P\bigl(y|\po(x)\bigr)
.
\end{equation}
If, on the other hand, the conditional
probabilities $P(z|\mbox{age})$ and $P^*(z|\mbox{age})$ are the same
in both cities,
and the inequality $P(z) \neq P^*(z)$ reflects genuine age differences,
equation (\ref{eq2}) is no longer valid, since the age difference may
be a critical factor in determining how people react to $X$.
We see, therefore, that the choice of the proper transport
formula depends on the causal context in which
population differences are embedded.

This example also demonstrates why
the invariance of $Z$-specific causal effects
should not be taken for granted. While
justified in Example~\ref{ex1}, with $Z$ = age,
it fails in Example~\ref{ex2}, in which $Z$ was
equated with ``language skills.'' Indeed,
using Figure~\ref{fig1}(b) for guidance, the
$Z$-specific effect of $X$ on $Y$ in NYC
is given by
\begin{eqnarray*}
&&P^*\bigl(y|\po(x), z\bigr)\\
&&\quad  = \sum_{{\mathrm{age}}} P^*\bigl(y|
\po(x), z, {\mbox{age}}\bigr) P^*\bigl({\mbox{age}}|\po(x), z\bigr)
\nonumber
\\
&&\quad = \sum_{{\mathrm{age}}} P^*\bigl(y|\po(x), {\mbox{age}}\bigr)
P^*({\mbox{age}}| z)
\nonumber
\\
&&\quad = \sum_{{\mathrm{age}}} P\bigl(y|\po(x), {\mbox{age}}\bigr)
P^*({\mbox{age}}| z).
\end{eqnarray*}
Thus, if the two populations differ in the
relation between age and skill, that is,
\[
P(\mbox{age}| z) \neq P^*(\mbox{age}| z)
\]
the skill-specific causal effect would differ as well.

The intuition is clear. A NYC person at
skill level $Z=z$ is likely to be in a totally different
age group from his skill-equals in Los Angeles and, since
it is age, not skill that shapes the way individuals
respond to treatment, it is only reasonable that
Los Angeles residents would respond differently
to treatment than their NYC counterparts at the very same skill level.

The essential difference between Examples \ref{ex1}
and \ref{ex2} is
that age is normally taken to be an exogenous variable
(not assigned by other factors in the model) while skills
may be indicative of earlier factors (age,
education, ethnicity) capable of modifying the causal
effect. Therefore, conditional on skill, the effect may
be different in the two populations.
%
%% insert-3 Example~3

%
%ex3 #&#
\begin{example}\label{ex3}
Examine the case where $Z$ is a $X$-dependent variable,
say a disease bio-marker, standing on the causal pathways
between $X$ and $Y$ as shown in Figure~\ref{fig1}(c).
Assume further that the disparity $P(z | x) \neq P^*(z | x)$ is
discovered and that, again, both the average and the
$z$-specific causal effect
$P(y|\po(x),z)$ are estimated in the LA experiment, for all levels of
$X$ and $Z$.
Can we, based on information given, estimate
the average (or $z$-specific) causal effect in the
target population of NYC?
%
%on ``surrogate endpoint'' %% \citet{prentice:89} %%(Prentice, 1989)
%% from 363=20
%{prentice:89,freedman:etal92,frangakis:rub02,baker:06,joffe:gre09,jp:bio11},
%that is,
%using the effect of $X$ on $Z$ to predict
%the effect of $X$ on $Y$ in a population with potentially
%differing characteristics. A robust solution
%to this problem is offered in \cite{pearl:bar10-r372}.\label{foot2}}
%
\end{example}

Here, equation (\ref{eq1}) is wrong because the overall causal effect
(in both LA and NYC) is no longer a simple average of the $z$-specific
causal effects. The correct weighing rule is
%
%e3.3 #&#
\begin{eqnarray}\label{eq3}
&&P^*\bigl(y|\po(x)\bigr) \nonumber\\[-8pt]\\[-8pt]
&&\quad  = \sum_z P^*\bigl(y|
\po(x), z\bigr) P^*\bigl(z|\po(x)\bigr),\nonumber %% (3)
\end{eqnarray}
which reduces to (\ref{eq1}) only in the special case where $Z$ is
unaffected by $X$. Equation (\ref{eq2}) is also wrong because we can
no longer argue, as we did in Example~\ref{ex2}, that $Z$ does not
affect $Y$, hence it can be ignored. Here, $Z$ lies on the causal
pathway between $X$ and $Y$ so, clearly, it affects their relationship.
What then is the correct transport formula for this scenario?

To cast this example in a more realistic setting,
let us assume that we wish to use $Z$ as a ``surrogate
endpoint'' to predict the efficacy of treatment $X$
on outcome $Y$, where $Y$ is too difficult and/or
expensive to measure routinely (\cite{prentice:89}; \cite{ellenberg:ham89}).
Thus, instead of considering
experimental and observational studies conducted at
two different locations, we consider two such studies
taking place at the same location, but at different times.
In the first study, we measure $P(y,z|\po(x))$ and discover that
$Z$ is a good surrogate, namely, knowing the effect of
treatment on $Z$ allows prediction of the effect of treatment
on the more clinically relevant outcome ($Y$) (\cite{joffe:gre09}).
Once $Z$ is proclaimed a ``surrogate endpoint,''
it invites efforts to find direct means of controlling $Z$.
For example, if cholesterol level is found to be a predictor
of heart diseases in a long-run trial, drug manufacturers would
rush to offer cholesterol-reducing substances for public
consumption. As a result, both the prior
$P(z)$ and the treatment-dependent probability
$P(z|\po(x))$ would undergo a change,
resulting in $P^*(z)$ and $P^*(z|\po(x))$, respectively.

We now wish to reassess the effect of the drug
$P^*(y|\po(x))$ in the new population and do it
in the cheapest possible way, namely, by conducting
an observational study to estimate $P^*(z,x)$,
acknowledging that confounding exists
between $X$ and $Y$ and that the drug affects $Y$ both
directly and through $Z$, as shown in Figure~\ref{fig1}(c).

Using a graphical representation to encode the
assumptions articulated thus far, and
further assuming that the disparity observed
stems only from a difference in people's
susceptibility to $X$ (and not due to a change in some unobservable
confounder), we will prove in Section~\ref{sec4}
that the correct transport formula should be
%
%e3.4 #&#
\begin{eqnarray}
\label{eq35} \hspace*{15pt}P^*\bigl(y|\po(x)\bigr)  = \sum_z
P\bigl(y|\po(x),z\bigr) P^*(z|x),
\end{eqnarray}
which is different from both (\ref{eq1}) and (\ref{eq2}).
It calls instead for the $z$-specific effects
to be reweighted by the conditional probability $P^*(z|x)$,
estimated in the target population.\footnote{Quite often the
possibility of running
a second randomized experiment to estimate $P^*(z|\po(x))$
is also available to investigators, though at
a higher cost. In such cases, a transport formula would be
derivable under more relaxed assumptions, for example, allowing for $X$
and $Z$ to be confounded.}
%The analysis of such transfers, from experimental to semi-experimental
%regimes takes us outside the scope of this paper \citet{bp:13a}.

To see how the transportability problem fits into the general scheme of
causal analysis discussed in Section~\ref{sec:21} (Figure~\ref{atlanta-slide1}), we note that, in our case, the data comes from two
sources, experimental (from the study) and nonexperimental (from the
target), assumptions are encoded in the form of selection diagrams, and
the query stands for the causal effect (e.g., $P^*(y|\po(x))$). Although
this paper does not discuss the goodness-of-fit problem, standard
methods are available for testing the compatibility of the selection
diagram with the data available.

%s4 #&#
\section{Formalizing Transportability} \label{sec3} %%Section~3:

%s4.1 #&#
\subsection{Selection Diagrams and Selection Variables}
\label{ss3.1}

The pattern that emerges from the examples discussed in Section~\ref{sec2} indicates that transportability is a causal, not statistical
notion. In other words, the conditions that license
transport as well as the formulas through which results are
transported depend on the causal relations between the variables
in the domain, not merely on their statistics.
For instance, it was important in Example~\ref{ex3} to ascertain that
the change in $P(z | x)$ was due to the change in the way $Z$ is
affected by $X$, but not due to a change in confounding conditions
between the two. This cannot be determined solely by comparing $P(z|x)$
and $P^*(z|x)$. If $X$ and $Z$ are confounded [e.g., Figure~\ref{fig4a}(e)], it is quite possible for the inequality $P(z|x) \neq
P^*(z|x)$ to hold, reflecting differences in confounding, while the way
that $Z$ is affected by $X$ (i.e., $P(z|\po(x))$) is the same in the two
populations---a different transport formula will then emerge for this case.
%
%f4 #&#
\begin{figure*}

\includegraphics{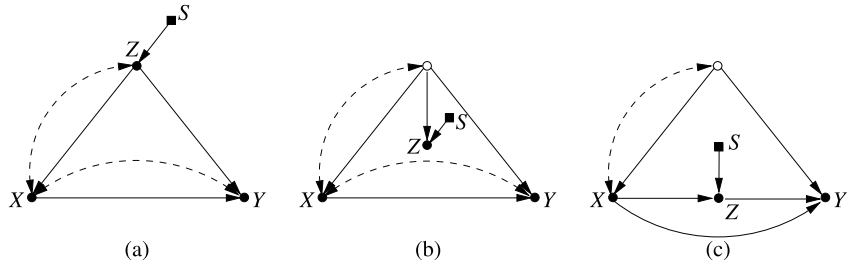}

\caption{Selection diagrams depicting specific versions of Examples
\protect\ref{ex1}--\protect\ref{ex3}.
In \textup{(a)}, the two populations differ in age distributions.
In \textup{(b)}, the populations differs in how $Z$ depends on age
(an unmeasured variable, represented by the hollow circle) and the age
distributions are
the same. In \textup{(c)}, the populations differ in how $Z$
depends on $X$. In all diagrams, dashed arcs (e.g., $X \dashleftarrow
\dashrightarrow Y$) represent the presence of latent variables
affecting both $X$ and $Y$.}
\label{fig2}
\end{figure*}

Consequently, licensing transportability requires knowledge of the
mechanisms, or processes, through which population differences come
about; different localization of these mechanisms yield different
transport formulae. This can be seen most vividly in Example~\ref{ex2}
[Figure~\ref{fig1}(b)] where we reasoned that
no reweighing is necessary
if the disparity $P(z) \neq P^*(z)$ originates with
the way language proficiency depends on age,
while the age distribution itself
remains the same.
Yet, because age is not measured,
this condition cannot be detected in
the probability distribution $P$,
and cannot be distinguished from an
alternative condition,
\[
P(\mbox{age}) \neq P^*(\mbox{age}) \quad \mbox{and}\quad  P(z|\mbox{age}) = P^*(z|
\mbox{age}),
\]
one that may require reweighting according
to equation (\ref{eq1}). In other words, every probability
distribution $P(x,y,z)$ that is compatible with the process of
Figure~\ref{fig1}(b) is also compatible with that of Figure~\ref{fig1}(a)
and, yet, the two processes dictate different transport formulas.

Based on these observations, it is clear that
if we are to represent formally the differences
between populations (similarly, between experimental settings
or environments), we must resort to a representation
in which the causal mechanisms are explicitly encoded
and in which differences in populations are
represented as local modifications of those
mechanisms.

To this end, we will use causal diagrams
augmented with a set, $S$, of
``selection variables,'' where each member of
$S$ corresponds to a mechanism by which
the two populations differ, and switching
between the two populations will be represented by
conditioning on different values of these $S$ variables.\footnote
{Disparities among populations or subpopulations can also arise from
differences in design; for example, if two samples are drawn by
different criteria from a given population. The problem of generalizing
between two such subpopulations is usually called \textit{sampling
selection bias} (\cite
{heckman:79}; \cite{hernan:etal04}; \cite{cole:stu10}; \cite{pearl:13-r409}; \cite{btp:14a}). In this
paper, we deal only with nature-induced, not man-made disparities. }

Intuitively, if $P(v|\po(x))$ stands for the distribution of a set $V$
of variables
in the experimental study (with X randomized)
then we designate by $P^*(v|\po(x))$
the distribution of $V$ if we were to
conduct the study on population $\Pi^*$ instead
of $\Pi$. We now attribute the difference between
the two to the action of a set $S$ of selection
variables, and write\footnote{Alternatively, one can represent the two
populations' distributions by $P(v| \po(x), s)$, and $P(v| \po(x), s^*)$,
respectively. The results, however, will be the same, since only the
location of $S$ enters the analysis.}\tsup{,}\footnote{Pearl (\citeyear{pearl:93h,pearl:09}, page
71), \citet{spirtes:etal93} and \citet
{dawid:02}, for example, use conditioning on auxiliary variables to
switch between experimental and observational studies. \citet
{dawid:02} further uses such variables to represent changes in
parameters of probability distributions.}
\[
P^*\bigl(v|\po(x)\bigr) = P\bigl(v| \po(x), s^*\bigr).
\]

The selection variables in $S$ may represent all factors by which
populations may differ or
that may ``threaten'' the transport of conclusions between populations.
For example, in Figure~\ref{fig2}(a) the age disparity $P(z) \neq P^*(z)$
discussed in Example~\ref{ex1} will be represented
by the inequality
\[
P(z) \neq P(z|s),
\]
where $S$ stands for all factors responsible for
drawing subjects at age $Z=z$ to NYC rather than
LA.

Of equal importance is the absence of an $S$ variable pointing
to $Y$ in Figure~\ref{fig2}(a), which encodes the assumption that
age-specific effects are invariant across the two populations.

%by an inequality of conditional probabilities \[ P(z|s_1) \neq
%P(z|s_1^*) \] where $S_1$ stands for all factors responsible for
%drawing subjects at age $Z=z$ to the corresponding populations.

This graphical representation, which we will call
``selection diagrams'' is defined as follows:\footnote{The assumption
that there are no structural changes between domains can be relaxed
starting with $D = G^*$ and adding $S$-nodes following the same
procedure as in Definition~\ref{def:sd}, while enforcing acyclicity.
In extreme cases in which the two domains differ in causal
directionality (\cite{spirtes:etal00}, pages 298--299), acyclicity cannot
be maintained. This complication as well as one created when $G$ is a
edge-super set of $G^*$
require a more elaborated graphical representation and lie beyond the
scope of this paper. }
%a more elaborated graphical representation and lie beyond the simple
%analysis introduced in this paper. }

%de4 #&#
\begin{definition}[(Selection diagram)] \label{def:sd}
Let $\langle M, M^* \rangle$ be a pair of structural causal models
(Definition~\ref{def11}) relative to domains
$\langle\Pi, \Pi^* \rangle$, sharing a causal diagram $G$.
$\langle M, M^* \rangle$ is said to induce a selection diagram $D$
if $D$ is constructed as follows:
\begin{enumerate}[2.]
\item Every edge in $G$ is also an edge in $D$.
\item$D$ contains an extra edge $S_i \rightarrow V_i$ whenever there
might exist
a discrepancy $f_i \neq f_i^*$ or $P(U_i) \neq P^*(U_i)$ between $M$
and $M^*$.
\end{enumerate}
\end{definition}

In summary, the $S$-variables locate the \textit{mechanisms} where
structural discrepancies between the two populations are suspected to
take place. Alternatively, the absence of a selection node pointing to
a variable represents the assumption that the mechanism responsible for
assigning value to that variable is the same in the two populations. In
the extreme case, we could add selection nodes to all variables,
which means that we have no reason to believe that the populations
share any mechanism in common, and this, of course would
inhibit any exchange of information among the populations. The
invariance assumptions between populations, as we will see,
will open the door for the transport of some experimental findings.

For clarity, we will represent the $S$ variables by
squares, as in Figure~\ref{fig2}, which uses selection
diagrams to encode the three
examples discussed in Section~\ref{sec2}. (Besides the $S$ variables,
these graphs also include additional latent variables, represented by
bidirected edges, which makes the examples more realistic.)
In particular, Figures~\ref{fig2}(a) and \ref{fig2}(b) represent,
respectively,
two different mechanisms responsible for the
observed disparity $P(z) \neq P^*(z)$. The first
[Figure~\ref{fig2}(a)] dictates transport formula (\ref{eq1}), while
the second [Figure~\ref{fig2}(b)] calls for direct, unadjusted
transport (\ref{eq2}). This difference stems from the location of the $S$
variables in the two diagrams. In Figure~\ref{fig2}(a), the
$S$ variable represents unspecified factors that
cause age differences between the two populations,
while in Figure~\ref{fig2}(b), $S$ represents
factors that cause differences in reading skills ($Z$)
while the age distribution itself (unobserved)
remains the same.

In this paper, we will address the issue of transportability
assuming that scientific knowledge about invariance
of certain mechanisms is available and encoded in the selection diagram
through the $S$ nodes. Such knowledge
is, admittedly, more demanding than that which shapes
the structure of each causal diagram in isolation. It is,
however, a prerequisite for any attempt to justify transfer of findings
across populations, which makes selection diagrams a mathematical
object worthy of analysis.

%s4.2 #&#
\subsection{Transportability: Definitions and Examples}
\label{ss3.2}

Using selection diagrams as the basic representational
language, and harnessing the concepts of
intervention, do-calculus, and identifiability (Section~\ref{sec2a}),
we can now give the
notion of transportability a formal definition.

%de5 #&#
\begin{definition}[(Transportability)] \label{def:transportability}
\label{def1a}
Let $D$ be a selection diagram relative to domains $\langle\Pi, \Pi
^*\rangle$. Let $\langle P, I \rangle$ be the pair of observational
and interventional distributions of $\Pi$, and $P^*$ be the
observational distribution of $\Pi^*$. The causal relation $R(\Pi^*)
= P^*(y | \po(x), z)$ is said to be transportable from $\Pi$ to $\Pi
^*$ in $D$ if $R(\Pi^*)$ is uniquely computable from $P, P^*, I$ in
any model that induces $D$.
\end{definition}

Two interesting connections between identifiability and
transportability are worth noting. First, note that all identifiable
causal relations in $D$ are also transportable, because they can be
computed directly from $P^*$ and require no experimental information
from $\Pi$. Second, note that given causal diagram $G$, one can
produce a selection diagram $D$ such that identifiability in $G$ is
equivalent to transportability in $D$. First set $D = G$, and then add
selection nodes pointing to all variables in $D$, which represents that
the target domain does not share any mechanism with its counterpart---this is equivalent to the problem of identifiability because the only
way to achieve transportability is to identify $R$ from scratch in the
target population.

While the problems of identifiability and transportability are related,
proofs of nontransportability are more involved than those of
nonidentifiability for they require one to demonstrate the nonexistence
of two competing models compatible with $D$, agreeing on $\{P, P^*, I\}
$, and disagreeing on $R(\Pi^*)$.

Definition~\ref{def1a} is declarative, and does not
offer an effective method of demonstrating transportability
even in simple models. Theorem~\ref{lem1} offers such a
method using a sequence of derivations in do-calculus.

%
%th1 #&#
\begin{theorem}\label{lem1} %%Lemma~4
Let $D$ be the selection diagram characterizing
two populations, $\Pi$ and $\Pi^*$, and $S$ a set of
selection variables in $D$.
The relation R = $P^*(y|\po(x), z)$ is transportable from
$\Pi$ to $\Pi^*$ if the expression $P(y|\po(x),z,s)$ is
reducible, using the rules of do-calculus, to an expression
in which $S$ appears only as a conditioning variable in do-free
terms.  %\hfill$\Box$
\end{theorem}

\begin{pf}
% \textbf{Proof :} \\
Every relation satisfying the condition of
Theorem~\ref{lem1} can be written as an algebraic combination
of two kinds of terms, those that involve $S$ and those
that do not. The former can be written as $P^*$-terms
and are estimable, therefore, from observations on $\Pi^*$,
as required by Definition~\ref{def1a}. All other terms, especially
those involving do-operators, do not contain $S$; they
are experimentally identifiable therefore in $\Pi$.
\end{pf}
%
%This criterion was proven to be both sufficient and necessary for
%causal effects, namely $R = P(y | do(x))$ \citet{bp:12b}. Theorem
%leading to the needed reduction when such a sequence exists.
This criterion was proven to be both sufficient and necessary for
causal effects, namely $R = P^*(y | \po(x))$ (\cite{bp:12b}). Theorem~\ref{lem1}, though procedural, does not specify the sequence of rules
leading to the needed reduction when such a sequence exists.
\citet{bp:13c} derived a complete procedural solution for this, based
on graphical method developed in (\cite{tian:pea02}; \cite{shpitser:pea06-r329}). Despite its completeness, however,
the procedural solution is not trivial, and we take here an alternative
route to establish a simple and transparent procedure for confirming
transportability, guided by two recognizable subgoals.

%de6 #&#
\begin{definition}[(Trivial transportability)]\label{def3}
A caus\-al relation $R$ is said to be \emph{trivially transportable} from
$\Pi$ to $\Pi^*$, if $R(\Pi^*)$ is identifiable from $(G^*, P^*)$.
\end{definition}

This criterion amounts to an ordinary test of identifiability of causal
relations using graphs, as given by Definition~\ref
{def-identifiability}. It permits us to estimate $R(\Pi^*)$ directly
from observational
studies on $\Pi^*$, unaided by causal information from $\Pi$.

%
%ex4 #&#
\begin{example}\label{ex7}
Let $R$ be the causal effect $P^*(y|\allowbreak \po(x))$
and let the selection diagram of $\Pi$ and $\Pi^*$
be given by $X \rightarrow Y \leftarrow S$, then $R$ is trivially
transportable, since $R(\Pi^*) = P^*(y|x)$.
\end{example}

Another special case of transportability occurs when a causal relation
has identical form in both domains---no recalibration is needed.

%de7 #&#
\begin{definition}[(Direct transportability)]\label{def2}
A causal relation $R$ is said to be \emph{directly transportable} from
$\Pi$ to $\Pi^*$, if $R(\Pi^*) = R(\Pi)$.
\end{definition}

A graphical test for direct transportability of $R = P^*(y|\po(x), z)$
follows from do-calculus and reads: $(S \ci Y |X,Z)_{G_{\overline
{X}}}$; in words, $X$ blocks all paths from $S$ to $Y$ once we remove
all arrows pointing to $X$ and condition on $Z$.
As a concrete example, this test is satisfied in Figure~\ref{fig2}(a)
and, therefore, the $z$-specific effects is the same in both
populations; it is directly transportable.

\begin{Remark*}
The notion of ``external validity'' as
defined by \citet{manski:07} %%Mansky (2007)
(footnote \ref{foot1}) corresponds
to Direct Transportability, for it requires that
$R$ retains its validity without adjustment,
as in equation (\ref{eq2}).
Such conditions preclude the use of information from $\Pi^*$ to
recalibrate $R$.
\end{Remark*}

%ex5 #&#
\begin{example}\label{ex4}
Let $R$ be the causal effect of $X$ on $Y$, and
let $D$ have a single $S$ node pointing to $X$, then
$R$ is directly transportable, because causal effects are
independent of the selection mechanism (see \cite{pearl:09}, pages 72 and 73).
\end{example}

%
%ex6 #&#
\begin{example}\label{ex6}
Let $R$ be the $z$-specific causal effect of $X$ on $Y$
$P^*(y|\po(x), z)$ where $Z$ is a set of
variables, and $P$ and $P^*$ differ only in the conditional
probabilities $P(z|\pa(Z))$ and $P^*(z|\pa(Z))$
such that $(Z \compos Y | \pa(Z))$, as shown in Figure~\ref{fig2}(b).
Under these conditions, $R$ is not directly transportable.
However, the $\pa(Z)$-specific causal effects
$P^*(y|\po(x), \pa(Z))$ are directly transportable,
and so is $P^*(y|\po(x))$.
Note that, due to the confounding arcs,
none of these quantities is identifiable.
\end{example}

%
%Let $R$ be the causal effect $P(y|do(x))$
%and let the selection diagram of $\Pi$ and $\Pi^*$
%be given by $X \rightarrow Y \leftarrow S$, with $X$ and $Y$
%confounded as shown in Fig.~\ref{fig4a}(b),
%then $R$ is not transportable, because $P^*(y|do(x)) = P(y|do(x),s)$
%cannot be reduced to a $s$-free expression
%using the rules of $do$-calculus.
%This is the smallest graph for which the causal effect
%is non-transportable.

%s5 #&#
\section{Transportability of causal effects---A graphical criterion}
\label{sec4}

We now state and prove two theorems that permit us
to decide algorithmically, given a selection diagram,
whether a relation is transportable between two
populations, and what the transport formula should be.

%th2 #&#
\begin{theorem}\label{th1}
Let $D$ be the selection diagram characterizing
two populations, $\Pi$ and $\Pi^*$, and $S$ the set of
selection variables in $D$.
The strata-specific causal effect
$P^*(y|\po(x),z)$ is transportable from
$\Pi$ to $\Pi^*$ if $Z$ d-separates $Y$ from $S$ in
the $X$-manipulated version of $D$, that is,
$Z$ satisfies $(Y\compos S|Z, X)_{D_{\overline{X}}}$.
%$\Box$
%kaoru, X^ should be X over-bar
\end{theorem}

\begin{pf}
\[
P^*\bigl(y|\po(x),z\bigr) = P\bigl(y|\po(x),z,s^*\bigr).
\]
From Rule \ref{r1} of do-calculus we have:
$P(y|\po(x),z,\allowbreak s^*) = P(y|\po(x), z)$ whenever
$Z$ satisfies $(Y \compos S|Z,\allowbreak  X)$ in $D_{\overline{X}}$.
This proves Theorem~\ref{th1}.
\end{pf}
%
% \textbf{Proof:} \\
%From Rule-1 of $\po$-calculus \citet[page 85]{pearl:09} %%(Pearl 2009, pp
%85)
%we have:
%$P(y|\po(x),z,s) = P(y|\po(x), z)$ whenever
%$Z$ satisfies $(Y \compos S|Z)$ in $D_{\overline{X}}$.
%This proves Theorem~\ref{th1}.

%
%de8 #&#
\begin{definition}[($S$-admissibility)]
A set $T$ of variables satisfying $(Y \compos S|T, X)$ in $D_{\overline
{X}}$ will be called
\mbox{$S$-admissible} (with respect to the causal effect of $X$~on~$Y$).
\end{definition}

%
%f5 #&#
\begin{figure}[b]

\includegraphics{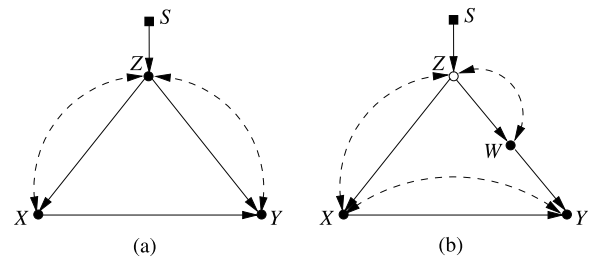}

\caption{Selection diagrams illustrating $S$-admissibility.
\textup{(a)} Has no $S$-admissible set while in \textup{(b)}, $W$ is
$S$-admissible.}
\label{fig3}
\end{figure}

%co1 #&#
\begin{corollary}\label{cor1}
The average causal effect $P^*(y|\allowbreak \po(x))$ is transportable from
$\Pi$ to $\Pi^*$ if there exists a set $Z$ of observed pretreatment covariates
that is \mbox{$S$-admissible}. Moreover, the transport formula is given
by the weighting of equation (\ref{eq1}). %$\Box$
\end{corollary}
%P^*(y|\po(x)) &= P(y|\po(x),s) \\
% &= \sum_z P(y|\po(x),z,s) P(z|\po(x),s) \\
% &= \sum_z P(y|\po(x),z) P(z|s)\\
% & \mbox{(using $S$-admissibility and Rule-3 of $\po$-calculus)}
% &= \sum_z P(y|\po(x),z) P^*(z) \label{eq10}

%
%ex7 #&#
\begin{example}\label{ex9}
The causal effect is transportable in Figure~\ref{fig2}(a), since
$Z$ is $S$-admissible, and in Figure~\ref{fig2}(b), where the
empty set is $S$-admissible. It is also transportable by the same
criterion in
Figure~\ref{fig3}(b), where $W$ is $S$-admissible, but not in
Figure~\ref{fig3}(a) where no $S$-admissible set exists.
\end{example}

\begin{figure*}

\includegraphics{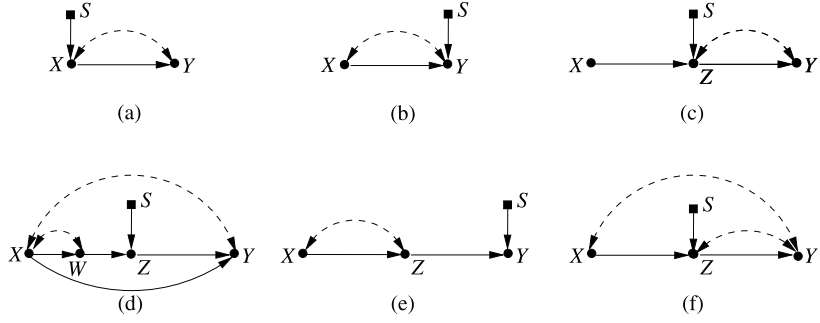}

\caption{Selection diagrams illustrating transportability. The causal
effect $P(y | \po(x))$ is (trivially) transportable in \textup{(c)} but not in
\textup{(b)} and \textup{(f)}. It is transportable in \textup{(a)}, \textup{(d)} and \textup{(e)} (see Corollary
\protect\ref{cor2}).}
\label{fig4a}
\end{figure*}

%
%co2 #&#
\begin{corollary}\label{cor2}  %%Corollary~2:
Any $S$ variable that is pointing directly into $X$
as in Figure~\ref{fig4a}(a), or that is d-separated from $Y$ in
$D_{\overline{X}}$ can be ignored.
%$\Box$
\end{corollary}

This follows from the fact that the empty set is
\mbox{$S$-admissible} relative to any such $S$ variable.
Conceptually, the corollary reflects the understanding that
differences in propensity to receive treatment do not hinder
the transportability of treatment effects; the randomization
used in the experimental study washes away such differences.

%
%f6 #&#

We now generalize Theorem~\ref{th1} to cases involving
treatment-dependent $Z$ variables, as in Figure~\ref{fig2}(c).

%th3 #&#
\begin{theorem}\label{th2} %%Theorem~2
%%The average causal effect $P(y|\po(x))$ is transportable from
%%$PP$ to $PP^*$ if either $P(y|\po(x))$ is identifiable in $PP^*$, or
%there exists a set of covariates, $Z$ (possibly affected by $X$) such
%that
%%\begin{enumerate}
%% \item$Z$ is $S$-admissible, and
%%%% \item$W$ satisfies $(X \compos Z|W,S)D_X$
%% \item$P(z|\po(x))$ is transportable.
%%\end{enumerate}
%%Revise Theorem~2
The average causal effect $P^*(y|\allowbreak \po(x))$ is transportable from
$\Pi$ to $\Pi^*$ if either one of the following conditions holds:
\begin{enumerate}[3.]
\item$P^*(y|\po(x))$ is trivially transportable.
\item There exists a set of covariates, $Z$ (possibly affected by $X$)
such that $Z$ is $S$-admissible and for which $P^*(z|\po(x))$ is transportable.
\item There exists a set of covariates, $W$
that satisfy $(X \ci Y|W)_{D_{\overline{X(W)}}}$ and for which
$P^*(w|\po(x))$ is transportable.
\end{enumerate}
%
%$ $* Optional. Try to further simplify each of the resultant factors
%through any $ $ standard identification procedure. \hfill%$\Box$
\end{theorem}

\begin{pf}
1. Condition 1 entails transportability.

2. If condition 2 holds, it implies
%
%e5.1 #&#
%e5.2 #&#
%e5.3 #&#
\begin{eqnarray}
&&P^*\bigl(y|\po(x)\bigr) \nonumber\\[-8pt]\\[-8pt]
&&\quad = P\bigl(y|\po(x),s\bigr)\nonumber
\\
&&\quad  = \sum_z P\bigl(y|\po(x),z,s\bigr) P\bigl(z|
\po(x),s\bigr)
\\
&&\quad  = \sum_z P\bigl(y|\po(x),z\bigr) P^*\bigl(z|
\po(x)\bigr).
\end{eqnarray}
We now note that the transportability of $P(z|\po(x))$ should reduce
$P^*(z|\po(x))$ to a star-free expression and would render
$P^*(y|\po(x))$ transportable.

3. If condition 3 holds, it implies
%
%e5.4 #&#
%e5.5 #&#
%e5.6 #&#
%e5.7 #&#
\begin{eqnarray}
&&P^*\bigl(y|\po(x)\bigr)\nonumber\\[-8pt]\\[-8pt]
&&\quad  =  P\bigl(y|\po(x),s\bigr)\nonumber
\\
&&\quad = \sum_w P\bigl(y|\po(x),w,s\bigr) P\bigl(w|
\po(x),s\bigr)
\\
&&\quad = \sum_w P(y|w,s) P^*\bigl(w|\po(x)\bigr)
\\
&&\qquad  \mbox{(by Rule \ref{r3} of do-calculus)}
\nonumber
\\
&&\quad =  \sum_w P^*(y|w) P^*\bigl(w|\po(x)\bigr).
\end{eqnarray}
We similarly note that the transportability of $P^*(w|\allowbreak \po(x))$
should reduce $P(w|\po(x),s)$ to a star-free expression and would render
$P^*(y|\po(x))$ transportable.
This proves Theorem~\ref{th2}.
\end{pf}

%
%ex8 #&#
\begin{example}
To illustrate the application of Theorem~\ref{th2}, let us apply it to
Figure~\ref{fig2}(c),
which corresponds to the surrogate endpoint problem discussed in
Section~\ref{sec2} (Example~\ref{ex3}).
Our goal is to estimate $P^*(y | \po(x))$---the effect of $X$ on $Y$ in
the new population created by changes in how $Z$ responds to $X$. The
structure of the problem permits us to satisfy condition 2 of Theorem \ref{th2}, since $Z$ is $S$-admissible and $P^*(z|\po(x))$ is trivially
transportable. The former can be seen from $(S \ci Y | X,
Z)_{G_{\overline{X}}}$, hence $P^*(y | \po(x), z) = P(y | \po(x),\break z))$;
the latter can be seen from the fact that $X$ and $Z$ and unconfounded,
hence $P^*(z|\po(x)) = P^*(z| x)$. Putting the two together, we get
%
%e5.8 #&#
\begin{equation}
%P^*(y | \po(x)) = \sum_z P^*(y | \po(x), z) P^*(z | x),
\hspace*{10pt}P^*\bigl(y | \po(x)\bigr) = \sum_z P
\bigl(y | \po(x), z\bigr) P^*(z | x),
\end{equation}
which proves equation (\ref{eq35}).
\end{example}

\begin{Remark*}
The test entailed by Theorem~\ref{th2} is recursive,
since the transportability of one causal effect
depends on that of another. However, given that
the diagram is finite and acyclic, the sets $Z$
and $W$ needed in conditions 2 and 3 of Theorem~\ref{th2}
would become closer and closer to $X$, and the iterative
process will terminate after a finite number of steps.
This occurs because the causal effects $P^*(z|\po(x))$
(likewise, $P^*(w|\po(x))$) is trivially transportable
and equals $P(z)$ for any $Z$ node that is not a descendant of $X$.
Thus, the need for reiteration applies only to those members
of $Z$ that lie on the causal pathways from $X$ to $Y$.
Note further that the analyst need not terminate the procedure upon
satisfying the conditions of Theorem~\ref{th2}. If one wishes to reduce
the number of experiments, it can continue until no further reduction
is feasible.
\end{Remark*}

%ex9 #&#
\begin{example} %%example 11
Figure~\ref{fig4a}(d) requires that we invoke both conditions of
Theorem~\ref{th2}, iteratively. To satisfy condition 2, we
note that $Z$ is $S$-admissible, and we need to
prove the transportability of $P^*(z|\po(x))$.
To do that, we invoke condition 3 and note that
$W$ \mbox{d-separates} $X$ from $Z$ in $D$. There remains
to confirm the transportability of $P^*(w|\po(x))$,
but this is guaranteed by the fact that the empty set
is $S$-admissible relative to $W$, since $(W \ci S)$. Hence, by
Theorem~\ref{th1} (replacing $Y$ with $W$) $P^*(w|\po(x))$ is transportable,
which bestows transportability on $P^*(y|\po(x))$.
Thus, the final transport formula (derived formally in the \hyperref[app1]{Appendix})
is:
%
%e5.9 #&#
\begin{eqnarray}\label{eq25}
&&P^*\bigl(y|\po(x)\bigr) \nonumber\\
&&\quad = \sum_z P\bigl(y|
\po(x),z\bigr) \\
&&\qquad {}\cdot
\sum_w P\bigl(w|\po(x)\bigr)
P^*(z|w). \nonumber
\end{eqnarray}
The first two factors of the expression are estimable in the experimental
study, and the third through observational studies on the target
population. Note that the joint effect $P(y,w,z|\po(x))$
need not be estimated in the experiment;
a decomposition that results in decrease of measurement cost and
sampling variability.
\end{example}

A similar analysis proves the transportability of the causal effect in
Figure~\ref{fig4a}(e) (see \cite{pearl:bar10-r372}).
The model of Figure~\ref{fig4a}(f), however, does not allow for the
transportability of $P^*(y| \po(x))$ as witnessed by the absence of
$S$-admissible set in the diagram, and the inapplicability of condition
3 of Theorem~\ref{th2}.

%
%ex10 #&#
\begin{example} %%example 12
\label{exx:10}
%% insert-11
To illustrate the power of Theorem~\ref{th2} in discerning
transportability and deriving transport formulae, Figure~\ref{fig5}
%
%f7 #&#
\begin{figure}[t]

\includegraphics{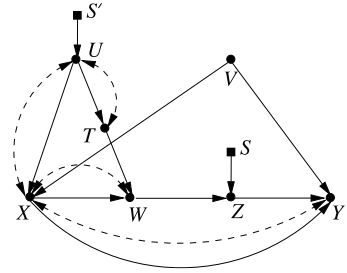}

\caption{Selection diagram in which the causal effect is shown to be
transportable in multiple iterations of Theorem \protect\ref{th2} (see the
\protect\hyperref[app1]{Appendix}).}
\label{fig5}
\end{figure}
represents a more intricate selection diagram, which
requires several iteration to discern transportability.
The transport formula for this diagram is given by (derived formally in the
\hyperref[app1]{Appendix}):
%
%e5.10 #&#
\begin{eqnarray}
\label{eq:xx} &&\hspace*{20pt}P^*\bigl(y | \po(x)\bigr)\nonumber \\
&&\hspace*{20pt}\quad = \sum_z
P\bigl(y|\po(x), z\bigr) \\
&&\hspace*{20pt}\qquad {}\cdot\sum_w P^*( z | w) \sum
_t P\bigl(w | \po(x), t\bigr) P^*(t).\nonumber
\end{eqnarray}
\end{example}

%% insert-14
The main power of this formula is
to guide investigators in deciding what
measurements need be taken in both
the experimental study and the target population.
It asserts, for example, that variables $U$ and $V$ need
not be measured. It likewise asserts that the $W$-specific
causal effects need not be estimated in the experimental
study and only the conditional probabilities
$P^*(z|w)$ and $P^*(t)$ need be estimated in the target population.
The derivation of this formulae is given in the \hyperref[app1]{Appendix}. %This result
%generalize the case where a simple adjustment, as given by eq. , is
%not sufficient 1

%%Insert as new paragraph at end of Section~4
Despite its power, Theorem~\ref{th2} in not complete,
namely, it is not guaranteed to approve all transportable
relations or to disapprove all nontransportable ones.
An example of the former is contrived in \citet{bp:12b}, where
an alternative, necessary and
sufficient condition is established in both graphical and algorithmic
form. Theorem~\ref{th2} provides, nevertheless,
a simple and powerful method of establishing transportability
in practice.

%s6 #&#
\section{Conclusions} \label{sec8}

%Informal discussions concerning the transportability
%of experimental results across populations have been
%going on for almost half a century, usually
%invoking the notions of ``external validity,'' ``heterogeneity''
%and others. The formalization offered in this paper
%embeds this discussion in a precise mathematical language,
%and provides researchers with the benefits of mathematical analysis,
%to improve the design and analysis of experimental
%studies.

Given judgements of how target populations may
differ from those under study, the paper offers a formal
representational language for making these
assessments precise and for deciding whether causal
relations in the target population
can be inferred from those obtained in an experimental study.
When such inference is possible,
the criteria provided by Theorems \ref{th1} and \ref{th2} yield transport
formulae, namely, principled
ways of calibrating the transported relations so as to
properly account for differences in the populations.
These transport formulae enable the investigator
to select the essential measurements in both the
experimental and observational
studies, and thus minimize measurement costs and sample
variability.

The inferences licensed by
Theorem~\ref{th1} and \ref{th2} represent worst case analysis, since
we have assumed,
in the tradition of nonparametric modeling, that every
variable may potentially be an effect-modifier (or moderator).
If one is willing to assume that certain relationships are
noninteractive, or monotonic as is the case in additive models,
then additional transport licenses may be issued,
beyond those sanctioned by Theorems \ref{th1} and~\ref{th2}.

While the results of this paper concern
the transfer of causal information
from experimental to observational studies,
the method can also benefit in transporting
statistical findings from one
observational study to another (\cite{pearl:bar10-r372}).
The rationale for such transfer is two-fold.
First, information from the first study may
enable researchers to avoid repeated
measurement of certain variables in the target
population.
Second, by pooling data from both
populations, we increase the precision in which
their commonalities are estimated and, indirectly,
also increase the precision by which the target
relationship is transported. Substantial
reduction in sampling variability can be thus
achieved through this decomposition (\cite{jp:r387}).

%``meta-analysis'' (\cite{glass:1976,hedges:etal85,owen:2009}),
Clearly, the same data-sharing philosophy can
be used to guide Meta-Analysis (\cite
{glass:1976}; \cite{hedges:etal85}; \cite{rosenthal:1995}; \cite{owen:2009}), where
one attempts to combine results from many experimental and
observational studies, each conducted on a different
population and under a different set of conditions, so
as to construct an aggregate measure of effect size that is
``better,'' in some formal sense, than any one study in isolation.
While traditional approaches aims to average out differences between studies,
our theory exploits the commonalities among the populations studied
and the target population. By pooling together commonalities and discarding
areas of disparity, we gain maximum use of the available samples (\cite{bp:13a}).

%The methodology described in this paper is also applicable in the
%selection of \textit{surrogate endpoints}, namely, variables that
%would allow good predictability of an outcome for both treatment and
%control. (\cite{ellenberg:ham89}) Using the representational power of
%``selection diagrams'', we have proposed a causally principled
%definition of ``surrogate endpoint'' and showed procedurally how valid
%surrogates can be identified in a complex network of cause-effect
%relationships (\cite{pearl:bar10-r372}.).

To be of immediate use,
our method relies on the assumption that the
analyst is in possession of sufficient background knowledge to
determine, at least qualitatively,
where two populations may differ from\vadjust{\goodbreak} one another.
This knowledge is not vastly different from that required in any
principled approach to causation in observational studies, since
judgement about possible effects of omitted factors is
crucial in any such analysis.
Whereas such knowledge may only be partially available, the analysis
presented in this paper is nevertheless essential for understanding
what knowledge is needed for the task to succeed
and how sensitive conclusions are to
knowledge that we do not possess.

Real-life situations will be marred, of course,
with additional complications that were not addressed
directly in this paper; for example, measurement errors, selection
bias, finite sample variability, uncertainty
about the graph structure and the possible existence
of unmeasured confounders between any two nodes in
the diagram. Such issues are not unique to transportability; they
plague any problem in causal analysis, regardless
of whether they are represented formally or ignored
by avoiding formalism. The methods
offered in this paper are representative of
what theory permits us to do in ideal situations,
and the graphical representation presented in this paper
makes the assumptions explicit and transparent.
Transparency is essential for reaching tentative
consensus among researchers and for
facilitating discussions to distinguish that which
is deemed plausible and important from that which is
negligible or implausible.

Finally, it is important to mention two recent extensions of the
results reported in this article. \citet{bp:13b} have addressed the
problem of transportability in cases where only a limited set of
experiments can be conducted at the source environment. Subsequently,
the results were generalized to the problem of
``meta-transportability,'' that is, pooling experimental results from
multiple and disparate sources to synthesize a consistent estimate of a
causal relation at yet another environment, potentially different from
each of the former (\cite{bp:13a}). It is shown that such synthesis
may be feasible from multiple sources even in cases where it is not
feasible from any one source in isolation.

%sA #&#
\begin{appendix}
\section*{Appendix} \label{app1}
Derivation of the transport formula for the causal effect in the model
of Figure~\ref{fig4a}(d) [equation (\ref{eq25})]:
%Derivation of the transport formula in the model of Fig.
\begin{eqnarray}
&&P^*\bigl(y|\po(x)\bigr) \nonumber\\
&&\quad = P\bigl(y|
\po(x),s\bigr)
\nonumber
\\
&&\quad =\sum_z P\bigl(y|\po(x),s, z\bigr) P\bigl(z |
\po(x), s\bigr)
\nonumber
\\
&&\quad = \sum_z P\bigl(y|\po(x), z\bigr) P\bigl(z |
\po(x), s\bigr)
\nonumber
\\
&&\qquad  \bigl(\mbox{2nd condition of Theorem~\ref{th2}},\nonumber\\
&&\qquad\hphantom{\bigl(} \mbox{$S$-admissibility of $Z$ of $CE(X,Y)$} \bigr)
\nonumber
\\
&&\quad = \sum_z P\bigl(y|\po(x), z\bigr) \nonumber\\
&&\qquad {}\cdot\sum
_w P\bigl(z | \po(x), w, s\bigr) P\bigl(w | \po(x), s\bigr)
\nonumber
\\
&&\quad = \sum_z P\bigl(y|\po(x), z\bigr)\nonumber\\
&&\qquad {}\cdot\sum
_w P(z | w, s) P\bigl(w | \po(x), s\bigr)
\\
&&\qquad  \bigl(\mbox{3rd condition of Theorem~\ref{th2}},\nonumber\\
&&\qquad \hphantom{\bigl(} (X \ci Z | W, S)_{D_{\overline{X(W)}}}
\bigr)
\nonumber
\\
&&\quad = \sum_z P\bigl(y|\po(x), z\bigr)\nonumber\\
&&\qquad {}\cdot \sum
_w P(z | w, s) P\bigl(w | \po(x)\bigr)
\nonumber
\\
&&\qquad  \bigl(\mbox{2nd condition of Theorem~\ref{th2}},\nonumber\\
&&\qquad \hphantom{\bigl(} \mbox{$S$-admissibility of the}\nonumber\\
&&\qquad \hphantom{\bigl(} \mbox{empty set $\{
\}$ of $CE(X,W)$} \bigr)
\nonumber
\\
&&\quad = \sum_z P\bigl(y|\po(x), z\bigr)\nonumber\\
&&\qquad {}\cdot \sum
_w P^*(z | w) P\bigl(w | \po(x)\bigr).\nonumber
\end{eqnarray}
Derivation of the transport formula for the causal effect in the model
of Figure~\ref{fig5} [equation (\ref{eq:xx})]:
%Derivation of the transport formula in the model of Fig.~\ref{fig5},
%(Eq. (\ref{eq:xx})).
%eA.2 #&#
\begin{eqnarray}
&&P^*\bigl(y|\po(x)\bigr)\nonumber \\
&&\quad = P\bigl(y|\po(x),s, s'\bigr) \nonumber\\
&&\quad = \sum
_z P\bigl(y|\po(x),s, s', z\bigr) P
\bigl(z | \po(x), s, s'\bigr)
\nonumber
\\
&&\quad = \sum_z P\bigl(y|\po(x), z\bigr) P\bigl( z |
\po(x), s, s'\bigr)
\nonumber
\\
&&\qquad  \bigl(\mbox{2nd condition of Theorem~\ref{th2}},\nonumber\\
&&\qquad \hphantom{\bigl(} \mbox{$S$-admissibility of $Z$ of $CE(X,Z)$}
\bigr)
\nonumber
\\
&&\quad =\sum_z P\bigl(y|\po(x), z\bigr) \nonumber\\
&&\qquad {}\cdot\sum
_w P\bigl( z | \po(x), s, s', w\bigr) P\bigl(w
| \po(x), s, s'\bigr)
\nonumber
\\
&&\quad = \sum_z P\bigl(y|\po(x), z\bigr)\nonumber\\
&&\qquad {}\cdot \sum
_w P\bigl( z | s, s', w\bigr) P\bigl(w |
\po(x), s, s'\bigr)
\nonumber
\\
&&\qquad  \bigl(\mbox{3rd condition of Theorem~\ref{th2}},\nonumber\\
&&\qquad \hphantom{\bigl(} \bigl(X \ci Z | W, S, S'
\bigr)_{D_{\overline{X(W)}}} \bigr)\nonumber
\\[-8pt]\\[-8pt]
&&\quad = \sum_z P\bigl(y|\po(x), z\bigr) \sum
_w P\bigl( z | s, s', w\bigr)\nonumber\\
&&\qquad {}\cdot \sum
_t P\bigl(w | \po(x), s, s', t\bigr) P\bigl(t
| \po(x), s, s'\bigr)
\nonumber
\\
&&\quad = \sum_z P\bigl(y|\po(x), z\bigr) \sum
_w P\bigl( z | s, s', w\bigr)\nonumber\\
&&\qquad {}\cdot \sum
_t P\bigl(w | \po(x), t\bigr) P\bigl(t | \po(x), s,
s'\bigr)
\nonumber
\\
&&\qquad  \bigl(\mbox{2nd condition of Theorem~\ref{th2}},\nonumber\\
&&\qquad \hphantom{\bigl(} \mbox{$S$-admissibility of $T$ on $CE(X,W)$}
\bigr)
\nonumber
\\
&&\quad = \sum_z P\bigl(y|\po(x), z\bigr) \sum
_w P\bigl( z | s, s', w\bigr) \nonumber\\
&&\qquad {}\cdot\sum
_t P\bigl(w | \po(x), t\bigr) P\bigl(t | s, s'
\bigr)
\nonumber
\\
&&\qquad  \bigl(\mbox{1st condition of Theorem~\ref{th2}/}\nonumber\\
&&\qquad \hphantom{\bigl(} \mbox{Rule \ref{r3} of do-calculus, $\bigl(X
\ci T | S, S'\bigr)_{D}$} \bigr)
\nonumber
\\
&&\quad = \sum_z P\bigl(y|\po(x), z\bigr) \sum
_w P^*( z | w) \nonumber\\
&&\qquad {}\cdot\sum_t P
\bigl(w | \po(x), t\bigr) P^*(t).\nonumber
\end{eqnarray}
\end{appendix}


\section*{Acknowledgments}
This paper benefited from discussions
with Onyebuchi Arah,
Stuart Baker,
Sander Greenland,\break
Michael Hoefler,
Marshall Joffe,
William Shadish,
Ian Shrier
and Dylan Small. We are grateful to two anonymous referees
for thorough reviews of this manuscript and for suggesting
a simplification in the transport formula of Example~\ref{exx:10}.
This research was supported in parts by NIH Grant \#1R01 LM009961-01,
NSF Grant \#IIS-0914211 and ONR Grant \#N000-14-09-1-0665.

%suskaldyti doi

% imsref loaded by jurgita.kaciuliene, 2014-07-09 13:42:04

\end{document}